\documentstyle[preprint,aps,eqsecnum]{revtex}

\def\beq#1{\begin{equation} \label{#1}}
\def\eeq{\end{equation}}

\input prepictex
\input pictex
\input postpictex
\newdimen\tdim
\tdim=\unitlength
\def\stpltsmbl{\setplotsymbol ({\small .})}

\newbox\phru
\setbox\phru=\hbox{\beginpicture
\setcoordinatesystem units <\tdim,\tdim>
\stpltsmbl
\setquadratic
\plot
0 0
2.5 3
5 0
7.5 -3
10 0
/
\endpicture}
\def\photonru #1 #2 *#3 /{\multiput {\copy\phru}  at
#1 #2 *#3 10 0 /}

\newbox\sru
\setbox\sru=\hbox{\beginpicture
\setcoordinatesystem units <\tdim,\tdim>
\stpltsmbl
\setquadratic
\plot
   0.0   0.0
   4.8   1.5
   7.5   5.0
   7.3   8.5
   5.0  10.0
   2.7   8.5
   2.5   5.0
   5.2   1.5
  10.0   0.0
/
\endpicture}
\def\springru #1 #2 *#3 /{\multiput {\copy\sru}  at
#1 #2 *#3 10 0 /}

\begin{document}
{
\tighten
\begin{center}
{\Large\bf Analysis of new charmless strange $B$ decay data leaves high
$B\rightarrow K\eta'$ and  $B\rightarrow K\eta' X$ still unexplained}  \\
\vrule height 2.5ex depth 0ex width 0pt
\vskip0.8cm
Harry J. Lipkin\,$^{b,c}$\footnote{e-mail: \tt ftlipkin@weizmann.ac.il} \\
\vskip0.8cm
{\it
$^b\;$School of Physics and Astronomy \\
Raymond and Beverly Sackler Faculty of Exact Sciences \\
Tel Aviv University, Tel Aviv, Israel\\
\vbox{\vskip 0.0truecm}
$^c\;$Department of Particle Physics \\
Weizmann Institute of Science, Rehovot 76100, Israel \\
and\\
High Energy Physics Division, Argonne National Laboratory \\
Argonne, IL 60439-4815, USA
} 
\end{center}

\vspace*{0.8cm}
\centerline{\bf Abstract}
\vspace*{4mm}

The question of whether the anomalously large $B \rightarrow K\eta' $ and  $B
\rightarrow K\eta'  X $ (inclusive) branching ratios are consistent with the
standard model or a sign of new physics is still open and in the  experimental
court. The nature of the extra hadron $X$ in $B \rightarrow K\eta'  X $ is
completely unknown. Directions for investigation and analysis of inclusive data
are suggested as well as for comparable inclusive data on $B \rightarrow K\eta
X $.

New data confirming the prediction  $BR(K^\pm \rho^o)  = BR(K^\pm \omega)$
support  flavor-topology $B \rightarrow VP$ sum rules based on the OZI rule and
emphasize the sharp contrast with the  failure of the analogous $B \rightarrow
PP $ sum rule due to the anomalously large $BR(B \rightarrow K\eta')$.
Confirmation of OZI validity in B-decay analyses for VP final states suggests
improving data on the analogous neutral decay difference  $BR(B^o \rightarrow
K^o \rho^o)  - BR(B^o \rightarrow K^o \omega)$ which measures tree-penguin
interference and possible direct CP violation.

A successful approximate isospin sum rule is rearranged and reinterpreted to
pinpoint tree-penguin interference in $B \rightarrow K \pi$ and $B \rightarrow
K \rho$ transitions. The magnitude of the interference is shown to still be at
the statistical noise level with present data.

\vfill\eject

\section{Two interesting sum rules and their implications}

\subsection{An interesting flavor-topology sum rule}

The large branching ratio 
$BR(B \rightarrow K \eta') \approx 70 \cdot 10^{-6}$ still remains a
problem\cite{etatranu}, together with the large inclusive branching ratio for 
$B \rightarrow K\eta'  X $ (inclusive).

The real problem here is that  $BR(B \rightarrow K \eta') \gg BR(B \rightarrow
K \pi)$. In the standard description of the $\eta$ and $\eta'$, their SU(3)
octet components  belong to the same pseudoscalar octet as the pions. Thus this
large difference  suggests a contribution to $BR(B \rightarrow K \eta')$ via
the SU(3) singlet component of  the $\eta$ and $\eta'$.

The necessity for this extra contribution is seen in the gross violation of the
sum rule\cite{bkpfsifin} which was derived specifically to test for such
contributions. 
\beq{sumrulepsbr} 
BR(B^\pm \rightarrow K^\pm \eta') + BR(B^\pm
\rightarrow K^\pm \eta) \leq  BR(B^\pm \rightarrow K^\pm \pi^o) +  BR(B^\pm
\rightarrow \tilde K^o \pi^\pm) 
\eeq 
where $\tilde K^o$ denotes $ K^o$ for the
$B^+$ decay and $\bar K^o$ for the $B^-$ decay. 
The experimental
values\cite{HFAG} in units of $10^{-6}$ are 
\beq{sumrulepsxbr} 
BR( K^\pm \eta')(70.8 \pm 3.4) + BR( K^\pm \eta)  (2.6 \pm 0.5) \leq 
BR( K^\pm \pi^o ) (12.1 \pm 0.8) +  BR( K^o \pi^\pm) (24.1 \pm 1.3) 
\eeq

    We review here the derivation which exploits known\cite{HJLCharm} 
flavor-topology\cite{CLOLIP12} characteristics of charmless strange  $B^\pm$
decays. Common calculations of weak decays are subject to  uncertainties
arising from unknown contributions of final state  interactions. The
flavor-topology approach automatically includes the  contributions to all
orders from all final state interactions described by quark-gluon  interactions
which conserve flavor SU(3).

The final states considered for $B^-$ decay all have the quark composition $s
\bar u q \bar q$ where $q \bar q$ denotes a pair of the same flavor which can
be $u \bar u$, $d \bar d$ or $s \bar s$. Charm admixture in the final state is
not considered.

We first review the flavor-topology properties of all the diagrams that 
can lead to final states $\bar K M$, where $\bar K$ denotes a $K^-$ or 
$\bar K^o$ or any analogous pair of $K^*$ resonances and $M$ denotes the 
members of any meson nonet, labeled $M_1$ for the unitary singlet state, 
$M_u$, $M_d$ and $M_s$ respectively for the $u\bar u$, $d\bar d$ and 
$s\bar s$ states and $M^-$ for the
$d \bar u$ state.

The $q \bar q$ pair observed in the final two-meson state may come from a very
complicated diagram involving many quarks and gluons. Flavor topology avoids
these  complications by focusing on the vertex in the diagram which creates 
this $q \bar q$ pair. There are only two possible vertices describing this 
pair creation, one where the pair is created from a gluon and one in which it
is created from a $W$ boson. The diagrams illustrated in figs. 1-7 show all
possible diagrams in which a $b \bar  u$ initial state enters a black box  and
emerges  as a state of a quark-antiquark pair and a boson, $W$ or gluon, that
hadronizes into the final state by QCD interactions including gluon exchanges
that do not change quark flavor quantum numbers and conserve flavor SU(3). The
black box  includes all the possible standard model diagrams with quark-gluon
interactions and flavor exchanges that conserve flavor SU(3).

The two relevant pair creation vertices are:

\begin{enumerate}

\item The pair is created by gluons and must therefore be a flavor singlet 
denoted by $(q \bar q)_1$. The four possible diagrams of this type are 
illustrated in figs. 1, 2, 3 and 4. The black boxes in these diagrams 
represent the sum of all possible diagrams which can lead from a $b \bar  u$
initial state to the $s \bar u G$ state. These diagrams include not only the 
gluonic penguin diagram and the annihilation diagram but also all tree diagrams
in which a $q \bar q$  pair is annihilated somewhere in the black box and
another pair is  created by interactions that conserve flavor-SU(3) symmetry.

The three contributions to the decay amplitude from the diagrams shown in 
figs. 2, 3 and 4 are equal by SU(3) symmetry.

\item There is no pair creation by gluons in the diagram and the  strange quark
must come from the initial weak vertex as an $s \bar u$  pair as illustrated in
figs. 5 and 6.  The black boxes in these diagrams  represent the sum of all
possible diagrams which can lead from a $b \bar  u$ initial state to the $u
\bar u W$ state. These diagrams include the tree diagrams along with all other
diagrams containing a tree and all strong rescatterings which do not change
the flavor of any quark. The line from the intitial spectator $\bar u$ antiquark
then passes theough the black box with gluon exchanges but no flavor change and 
combines either with either of the two final quarks to make either a $K^-$  or 
or an $M_u$.  

These two contributions to the decay amplitude add coherently and are 
commonly called color favored and color suppressed tree diagrams.

\end{enumerate}

Figs. 1-7 show all the flavor topology diagrams considered here. The 
treatment is exact in the standard model except for the effects of 
flavor-SU(3) symetry breaking and the contributions from other diagrams 
like the electroweak penguin diagram.

We further assume the A...Z\cite{ALS,PATOZ} or 
OZI\cite{Okubo,Zweig,Iizuka} rule that the quark and antiquark in the 
flavor singlet pair created by gluons cannot both appear in the same 
hadron; i.e. that the gluonic hairpin diagram illustrated in fig. 1 is 
forbidden. Since the remaining diagrams satisfy flavor-SU(3) symmetry 
there are only two independent amplitudes for describing the creation of 
the $q \bar q$ pair in the final two-meson state.

The decays are therefore
described by three parameters:
\begin{enumerate}

\item The flavor singlet
$s \bar u (q \bar q)_1$ amplitude summing the contributions from 
diagrams illustrated in figs. 2, 3 and 4, in which the quark and antiquark 
appear in
different final mesons.

\item A $K^- M_u$ amplitude summing the contributions from diagrams 
illustrated in figs. 5 and 6.

\item A relative phase.
\end{enumerate}

For decays into two pseudoscalar mesons, the one relation obtainable
between the decays to four final states is the sum rule:
\beq{sumruleps}
\tilde \Gamma(B^\pm \rightarrow K^\pm \eta') + \tilde \Gamma(B^\pm
\rightarrow K^\pm \eta) \leq  \tilde \Gamma(B^\pm \rightarrow K^\pm \pi^o)
+  \tilde \Gamma(B^\pm \rightarrow \tilde K^o \pi^\pm)
\eeq 
where $\tilde \Gamma$ denotes the partial width when phase space corrections
due to mass differences are neglected. $\tilde K^o$ denotes $ K^o$ for the
$B^+$ decay and $\bar K^o$ for the $B^-$ decay.

The equality holds in the flavor-SU(3) limit. The direction of the inequality
follows from the assumption that SU(3) symmetry breaking will suppress the $s
\bar s$ contribution to the singlet $(q \bar q)_1$. The left hand side of the
sum rule (\ref{sumruleps}) is seen to be invariant under the $\eta - \eta'$
mixing transformation. Thus the result holds for the pseudoscalar nonet with
any $\eta - \eta'$ mixing angle.

This result can be expressed as an inequality in terms of branching ratios 
since the difference in the phase space factors between the left and right due
to the low pion mass depresses the branching ratios to the $K\eta$ and  and
$K\eta'$ more than those for the $K\pi$ final states. 
This gives the inequality sum rule (\ref{sumrulepsbr}).

The experimental violation of the pseudoscalar sum rule (\ref{sumrulepsbr})
indicates a violation of one of the basic assumptions leading to the
derivation. Since SU(3) symmetry breaking which keeps the pseudoscalar nonet
intact is not likely to produce such a result, the most likely explanations are
a breakdown of the pseudoscalar nonet picture or a violation of the OZI rule.
There have been suggestions of an additional contribution\cite{PengSU3} outside
the nonet  like a glueball, charm admixture or radial
excitation\cite{Hareta,FRIJACK,ATSON} in the $\eta'$ wave function or an A...Z
or OZI-violating hairpin diagram illustrated by fig. 1.  

It is therefore of interest to look for tests of the nonet picture. New direct
experimental tests of this standard mixing picture in $B$ decays to $\eta$ and
$\eta'$ together with charmonium  have been suggested\cite{nudat0111}. he
suggestion that  
the OZI rule is violated in these decays leads to expectations for similar
violations or additional contributions to decays to final states containing the
$\eta'$ to occur elsewhere.

However the present new data\cite{HFAG}  show no such OZI violation in any of 
the vector pseudoscalar decays and no additional $\eta'$ enhancement in  other
related decays; e.g. $ B^\pm \rightarrow K^\pm \eta$, $B^\pm \rightarrow
K^{*\pm} \eta'$, or $B^\pm \rightarrow K^{*\pm} \eta$.  Further tests are
obtainable by looking for $\eta'$ enhancement in  $B^\pm \rightarrow \pi^\pm
\eta'$, $B^\pm \rightarrow \pi^\pm \eta$,  $B^\pm \rightarrow \rho^\pm \eta'$
and $B^\pm \rightarrow \rho^\pm \eta$  decays, where a charm admixture can give
direct CP violation asymmetry.

A serious combined analysis of all decays involving $\eta$ and $\eta'$ final
states might show which decays conform to the conventional picture where the
$\eta$ and $\eta'$ behave as normal members of the pseudoscalar nonet and where
there is evidence of anomalous enhancement and a possible violation of the
standard mixing picture.  Here the open question of the nature of
the enhancement in the inclusive $B \rightarrow K \eta' X $ deserves a serious
experimental  investigation along with the analogous inclusive decays $B
\rightarrow K \eta X $. 

The paradox of the absence of strong anomalous enhancements elsewhere is
sharpened by the agreement of new data\cite{HFAG} with the similar
flavor-topology relations  holding for the vector-pseudoscalar final states.
There are two cases depending upon whether the strange or the nonstrange meson
is a vector. In the nonstrange vector case, the ideal $\omega -\phi$ mixing
separates the sum rule into two equalities\cite{bkpfsifin},

\beq{eqv1}
BR( K^\pm \rho^o)  = BR( K^\pm \omega)
\eeq

\beq{eqv2}
BR( K^\pm \phi)   \leq BR( K^o \rho^\pm )
\eeq 
where the $\rho - \omega$ approximate degeneracy preserves the approximate
equalities of the branching ratios in the relation (\ref{eqv1}).
The inequality in the relation (\ref{eqv2}) arises from the $\rho - \phi$
mass difference.

The new data\cite{HFAG} show agreement with experiment for the equality 
(\ref{eqv1}). 
\beq{sumrulepv1}
BR( K^\pm \rho^o) 
(5.15 \pm 0.9) =
BR( K^\pm \omega)
(5.1 \pm 0.7)
\eeq
Better statistics are needed for a significant test of the equality 
(\ref{eqv2}).
\beq{sumrulepv2}
BR( K^\pm \phi) 
(9.7 \pm 1.5) \leq
BR( K^o \rho^\pm )
(\leq 48 )
\eeq

The other vector-pseudoscalar sum rule with the strange vector is
\beq{sumrulev}
\tilde \Gamma(B^\pm \rightarrow K^{*\pm} \eta') + \tilde \Gamma(B^\pm
\rightarrow K^{*\pm} \eta) \leq  \tilde \Gamma(B^{*\pm} 
\rightarrow K^{*\pm}\pi^o)
+  \tilde \Gamma(B^\pm \rightarrow \tilde K^o \pi^\pm)
\eeq

The experimental values from the new data\cite{HFAG} also show no evidence for
a strong violation indicating a large contribution \cite{Hareta,FRIJACK,ATSON}
of the type needed to explain the large violation of the sum
rule(\ref{sumrulepsbr}).

\beq{sumrulevp}
BR( K^{*\pm} \eta') 
(\leq 14) +
BR(K^{*\pm} \eta) 
(24.3 \pm 3.0) \leq
BR(K^{*\pm}\pi^o) 
(6.9 \pm 2.3) +
BR( K^{*o} \pi^\pm)
(9.7 \pm 1.2)
\eeq

The success of the equality (\ref{eqv1}) and the absence of any strong violation
in the other relations raise the question of
why the additional contribution needed to explain the
violation required for the pseudoscalar sum rule (\ref{sumrulepsbr}) does not
appear elsewhere.

In these $B^\pm$ decays all amplitudes arising from the $b \rightarrow u \bar u
s$ transition depend only upon a single sum of the color-favored and
color-suppressed tree contributions illustrated respectively in figs. 5 and 6.
This simplification provides predictive power and allows crucial tests of the
basic assumptions for  charged $B$ decays. This simplification  does not arise
in the neutral decays, where independent color-favored and  color-suppressed
tree contributions must be separately considered. Thus amplitudes derived here
for charged decays are not simply related by isospin to amplitudes for neutral
decays.

The possible implication of these results for the neutral decays is discussed
below.

\subsection{An approximate isospin sum rule that agrees with experiment}

An approximate equality\cite{approxlip,approxgr} has been
expressed as the ``Lipkin sum rule"\cite{Ali}.

\beq{sumruleapp}
R_L \equiv 2{{\Gamma(B^+ \rightarrow K^+ \pi^o) + \Gamma(B^o
\rightarrow K^o \pi^o)} \over {\Gamma(B^+ \rightarrow K^o \pi^+ )
+  \Gamma(B^o
\rightarrow K^+ \pi^-)}} \approx 1
\eeq

However, writing the relation in this form obscures the fact that it has real
significance only if there is interference between the dominant penguin and
another amplitude leading to an I=3/2 final state. It is trivially satisfied if
the decays are described entirely by a pure penguin contribution. This
physics can be seen explicitly by  rearranging the sum rule (\ref{sumruleapp})
as the approximate equality, 
\beq{eqapp}
2BR(B^+ \rightarrow K^+ \pi^o) 
- BR(B^+ \rightarrow K^o \pi^+ ) \approx 
BR(B^o \rightarrow K^+ \pi^-)  - 2BR(B^o\rightarrow K^o \pi^o)
\eeq
where for simplicity we express the result in terms of  branching ratios, which
can be corrected for lifetime differences when there is sufficient precision.

For the case where the decays are described entirely by a pure penguin
contribution, the final states are both isospin eigenstates with I=1/2 and both
sides of the relation (\ref{eqapp}) vanish. In this case the relation
reduces to the trivial 0=0. For the case where there is an
additional small contribution leading to an I=3/2 final state, the approximate
equality (\ref{eqapp}) relates the contributions to the charged and neutral
decays from the  interference  between this I=3/2 final state and the dominant
penguin.

Substituting experimental values gives
\beq{expeqapp}
2\cdot (12.1\pm 0.8)
- (24.1 \pm 1.3) = 0.1 \pm 2.1 \approx 
(18.2 \pm 0.8) - 
2\cdot (11.5 \pm 1.0) = - 4.8 \pm 2.2
\eeq

The agreement here within two standard deviations also exhibits a finite small 
contribution of tree-penguin interference whose true value is unfortunately
down in the noise of the statistics. However, its smallness justifies the
approximation of treating the tree contribution only in first order.

The same approximate equalities can be examined for the vector-pseudoscalar
final states.
\beq{ksteqapp}
2BR(B^+ \rightarrow K^{*+} \pi^o) 
- BR(B^+ \rightarrow K^{*o} \pi^+ ) \approx 
BR(B^o \rightarrow K^{*+} \pi^-)  - 2BR(B^o\rightarrow K^{*o} \pi^o)
\eeq
Substituting experimental values gives
\beq{exkstapp}
2\cdot (6.9\pm 2.3)
- (9.2 \pm 1.2) = 4.6 \pm 4.8 \approx 
(12.1 \pm 1.8) - 
2\cdot (1.7 \pm 0.8) = 8.7 \pm 2.4
\eeq

Here again, the agreement is within statistics but the error is still too large
to give a significant estimate of the tree-penguin interference.

\subsection{Some implications for neutral decays and direct CP violation}

The success of the equality (\ref{eqv1}) between the  $K^\pm \rho^o$   and
$K^\pm \omega$ decays of the charged $B$'s has interesting implications for
these decays of neutral $B$'s. The equality (\ref{eqv1}) follows for charged
$B$ decays, where the initial state has a spectator $u$ quark, because both the 
$\rho^o$ and $\omega$ are produced via their $u \bar u$ component. There is
no diagram producing the $d \bar d$ component if the OZI rule holds. The success
of the equality (\ref{eqv1}) thus indicates that the OZI rule holds and that it
can be expected also to hold for the neutral decays.

This physics underlying the equality (\ref{eqv1}) is exactly the same as that
which motivated the originally surprising prediction\cite{ALS} of equality for
the strong interaction reactions  \beq{ALS} \sigma(K^- p\rightarrow \Lambda
\omega) =   \sigma(K^- p\rightarrow \Lambda \rho^o)   \end{equation} Here the
$\omega - \rho^o$ equality follows also because there is no diagram  producing
a $d \bar d$ component.

For the neutral $B_d$ decays, where the initial state has a spectator $d$
quark, the penguin diagram produces both the $\rho^o$ and $\omega$ via their $d
\bar d$ component, as in fig. 3 with the spectator antiquark changed to a 
$\bar d$. The color suppressed tree diagram still produces both via their  $u
\bar u$ component as in fig. 5. Note that the color favored tree diagram
produces only charged vector mesons in neutral $B_d$ decays as in fig. 4 with
the spectator antiquark changed to a $\bar d$. The tree-penguin interference
thus has opposite phase for  $K^o \rho^o$   and $K^o \omega$ decays of neutral
$B$'s. A difference between $BR( K^o \rho^o)$ and  $BR( K^o\omega)$ would
indicate the presence of  tree-penguin interference and therefore a possibility
for observing direct CP violation in the charged as well as the neutral $B
\rightarrow K \rho $ and $B \rightarrow K \omega$ decays.

It is therefore of interest to refine the data for the $K \rho^o$ and 
$ K\omega$ decays of both charged and neutral $B$'s to look for evidence for
tree-penguin interference.

\section{Directions for investigating inclusive $B \rightarrow K\eta' X $}

To gain a better understanding for why the additional contribution needed to
explain the large violation of the sum  rule(\ref{sumrulepsbr}) does not seem
to appear elsewhere, we return  to the large inclusive branching ratios which
also are not yet sufficiently investigated.

\subsection {Parity Selection Rules from Gluonic Penguin Diagrams for Final
States Containing the $\eta$ and $\eta'$}

Consider the model where the gluonic penguin diagram produces the $\eta$ and
$\eta'$  in charmless strange final states both via the $u \bar u$ (or $d \bar
d$) and $s \bar s$ components of these mesons. As shown\cite{etatranu} in figs.
2-4 and the equation under fig. 4, the amplitudes for the   two components
interfere constructively for the $\eta'$  and destructively for the $\eta$ in
all final states of even parity and vice versa for states of odd  parity.

This model predicts that the $\eta$ should appear in states strongly dominated
by ODD parity and the $\eta'$  in states of EVEN parity. This prediction should
be violated in most models which introduce some other mechanism for explaining
the large $\eta'$  enhancement found in the $K\eta'$ final state.

This leads to the large $\eta'$/$\eta$ ratio for the $K\eta$ and $K\eta'$ 
final states and the reverse\cite{etatranu} for the $K^*$(890)$\eta$ and 
$K^*\eta'$.

\beq{nonprob}
BR( K^\pm \eta') \gg BR( K^\pm \eta); ~ ~ ~ \rm{while} ~ ~ ~ 
BR( K^{*\pm}\eta')   \ll BR(K^{*\pm} \eta) 
\eeq

\subsection {Experimental consequences of the Parity Selection Rules}

We now list some further experimental consequences of this
parity rule\cite{etatranu} which can be checked possibly with already available data.

\begin{enumerate}

\item The $K\pi\eta$ and $K\pi\eta'$ states all have odd parity, even when
the $K\pi$ is not in a $K^*$. Therefore the model predicts that the $K\pi\eta$
should be much stronger than $K\pi\eta'$ when summed over all final states and
that this holds both for the charged and neutral B decays. Thus one can get
better statistics for a comparison of the $\eta$ to the $\eta'$  by summing
over all of them. If this gives a strong enhancement of the $\eta$ over the
$\eta'$  this can be strong evidence against models that produce the $\eta'$ 
via the SU(3) singlet component; e,g, gluons, anomaly or intrinsic charm, since
it is hard to find reasons why this should hold only for a two-pseudoscalar
state and not for a three-pseudoscalar state.

\item  If $B \rightarrow K\eta'  X $ (inclusive)
data are plotted against the mass of the state $X$ recoiling
against the $\eta'$, the parity arguments suggest that the
contribution to the $K \eta' \pi$ final state should be small; rather
$K\eta\pi$ should be strong. This can be tested by noting whether
the $X$ mass in the strong signal observed for $K\eta'X$ is a pion mass or
more than $2 m_\pi$. If the single pion contribution is indeed small,
large contributions are ruled out from states\cite{HFAG} in which the state $X$ 
is the scalar resonance $K_o(1430)\rightarrow (93\%-K\pi)$
or the tensor $K_2(1430)\rightarrow (50\%-K\pi)$ and limit contributions from 
higher
states like $K^*(1680)\rightarrow (39\%-K\pi)$. This test would allow ruling out these
resonances without needing any complicated fits to mass plots and simplify the
analysis of what is left as well as putting constraints on models which explain
the $\eta'$ excess by some kind of singlet creation of the $\eta'$.

\item The measurement of the TRANSVERSITY in the final states $\eta \rho K$,
$\eta' \rho K$, $\eta \pi K^*(890)$ and $\eta' \pi K^*(890)$. This is the
measurement of the polarization of the vector meson in its rest frame with
respect to an axis normal to the VPP plane\cite{Trans}. This gives an
unambiguous signal for the PARITY of the final state (whether it is $0^+$ or
$0^-$) independent of the quantum numbers of the $K\pi\pi$ state recoiling
against the $\eta$ or $\eta'$

\item An $\eta$ or $\eta'$  recoiling against a $K^*$ resonance with NATURAL
parity (even P for even J and odd P for odd J) has odd parity and should give a
final state favoring the $\eta$ over the $\eta'$ . The opposite is true for a
recoil against a state with UNNATURAL parity. The $K$ and $K^*(890)$ states are
special cases of this prediction.

\item One should look for $K\eta$ and $K\eta'$ resonances in the states
$K\eta X$ and $K\eta' X$. Here the even parity resonances should favor the
$\eta'$  and the odd parity resonances favor the $\eta$.

\end{enumerate}

\section*{Acknowledgements}

This research was supported in part by the U.S. Department of Energy, Division
of High Energy Physics, Contract W-31-109-ENG-38. It is a pleasure to thank
Michael Gronau, Yuval Grossman, Zoltan Ligeti, Yosef Nir, Jonathan Rosner, J.G.
Smith, and  Frank Wuerthwein for discussions and comments.

%
\catcode`\@=11 
\def\references{
\ifpreprintsty \vskip 10ex
%
\hbox to\hsize{\hss \large \refname \hss }\else
\vskip 24pt \hrule width\hsize \relax \vskip 1.6cm \fi \list
{\@biblabel {\arabic {enumiv}}}
{\labelwidth \WidestRefLabelThusFar \labelsep 4pt \leftmargin \labelwidth
\advance \leftmargin \labelsep \ifdim \baselinestretch pt>1 pt
\parsep 4pt\relax \else \parsep 0pt\relax \fi \itemsep \parsep \usecounter
{enumiv}\let \p@enumiv \@empty \def \theenumiv {\arabic {enumiv}}}
\let \newblock \relax \sloppy
  \clubpenalty 4000\widowpenalty 4000 \sfcode `\.=1000\relax \ifpreprintsty
\else \small \fi}
\catcode`\@=12 

{\begin{figure}[htb]
$$\beginpicture
\setcoordinatesystem units <\tdim,\tdim>
\stpltsmbl
\putrule from -25 -30 to 50 -30
\putrule from -25 -30 to -25 30
\putrule from -25 30 to 50 30
\putrule from 50 -30 to 50 30
\plot -25 -20 -50 -20 /
\plot -25 20 -50 20 /
\plot 50 20 120 40 /
\plot 50 0 120 20 /
\springru 50 -20 *3 /
\plot 120 -20 90 -20 120 -40 /
\put {$b$} [b] at -50 25
\put {$\overline{u}$} [t] at -50 -25
\put {$s$} [l] at 125 40
\put {$\overline{u}$} [l] at 125 20
\put {$q$} [l] at 125 -20
\put {$\overline{q}$} [l] at 125 -40
\put {$\Biggr\}$ $K^-(\vec k)$}[l] at 135 30
\put {$\Biggr\}$ $M_1(-\vec k)$} [l] at 135 -30
\put {$G$} [t] at 70 -25
\setplotsymbol ({\large .})
\setshadegrid span <1.5\unitlength>
\hshade -30 -25 50 30 -25 50 /
\linethickness=0pt
\putrule from 0 0 to 0 60
\endpicture$$
\caption{\label{fig-1}} \hfill  ``Gluonic hairpin'' diagram. $G$
denotes any number of gluons. \hfill~
\end{figure}}

\pagebreak 
{\begin{figure}[htb]
$$\beginpicture
\setcoordinatesystem units <\tdim,\tdim>
\stpltsmbl
\putrule from -25 -30 to 50 -30
\putrule from -25 -30 to -25 30
\putrule from -25 30 to 50 30
\putrule from 50 -30 to 50 30
\plot -25 -20 -50 -20 /
\plot -25 20 -50 20 /
\plot 50 20 120 40 /
\plot 50 -20 120 -40 /
\springru 50 0 *3 /
\plot 120 20 90 0 120 -20 /
\put {$b$} [b] at -50 25
\put {$\overline{u}$} [t] at -50 -25
\put {$s$} [l] at 125 40
\put {$\overline{u}$} [l] at 125 20
\put {$u$} [l] at 125 -20
\put {$\overline{u}$} [l] at 125 -40
\put {$\Biggr\}$ $K^-(\vec k)$} [l] at 135 30
\put {$\Biggr\}$  $M_u(-\vec k)$} [l] at 135 -30
\put {$G$} [t] at 70 -5
\setshadegrid span <1.5\unitlength>
\hshade -30 -25 50 30 -25 50 /
\linethickness=0pt
\putrule from 0 0 to 0 60
\endpicture$$
\caption{\label{fig-2}} \hfill Strong $u \bar u$ pair creation. $G$ denotes any number of
gluons. \hfill~ \end{figure}}

{\begin{figure}[htb]
$$\beginpicture
\setcoordinatesystem units <\tdim,\tdim>
\stpltsmbl
\putrule from -25 -30 to 50 -30
\putrule from -25 -30 to -25 30
\putrule from -25 30 to 50 30
\putrule from 50 -30 to 50 30
\plot -25 -20 -50 -20 /
\plot -25 20 -50 20 /
\plot 50 20 120 40 /
\plot 50 -20 120 -40 /
\springru 50 0 *3 /
\plot 120 20 90 0 120 -20 /
\put {$b$} [b] at -50 25
\put {$\overline{u}$} [t] at -50 -25
\put {$s$} [l] at 125 40
\put {$\overline{d}$} [l] at 125 20
\put {$d$} [l] at 125 -20
\put {$\overline{u}$} [l] at 125 -40
\put {$\Biggr\}$ $\bar K^o(\vec k)$} [l] at 135 30
\put {$\Biggr\}$  $M^-(-\vec k)$} [l] at 135 -30
\put {$G$} [t] at 70 -5
\setshadegrid span <1.5\unitlength>
\hshade -30 -25 50 30 -25 50 /
\linethickness=0pt
\putrule from 0 0 to 0 60
\endpicture$$
\caption{\label{fig-3}} \hfill Strong $d \bar d$ pair creation. $G$ denotes any number of
gluons. \hfill~ \end{figure}}

{\begin{figure}[htb]
$$\beginpicture
\setcoordinatesystem units <\tdim,\tdim>
\stpltsmbl
\putrule from -25 -30 to 50 -30
\putrule from -25 -30 to -25 30
\putrule from -25 30 to 50 30
\putrule from 50 -30 to 50 30
\plot -25 -20 -50 -20 /
\plot -25 20 -50 20 /
\plot 50 20 120 40 /
\plot 50 -20 120 -40 /
\springru 50 0 *3 /
\plot 120 20 90 0 120 -20 /
\put {$b$} [b] at -50 25
\put {$\overline{u}$} [t] at -50 -25
\put {$s$} [l] at 125 40
\put {$\overline{s}$} [l] at 125 20
\put {$s$} [l] at 125 -20
\put {$\overline{u}$} [l] at 125 -40
\put {$\Biggr\}$  $M_s(\vec k)$}[l] at 135 30
\put {$\Biggr\}$  $K^-(-\vec k)$} [l] at 135 -30
\put {$G$} [t] at 70 -5
\setshadegrid span <1.5\unitlength>
\hshade -30 -25 50 30 -25 50 /
\linethickness=0pt
\putrule from 0 0 to 0 60
\endpicture$$
\caption{\label{fig-4}} \hfill Strong $s \bar s$ pair creation. $G$ denotes any number of
gluons. \hfill~ \end{figure}}

$$ A[M_s(\vec k)K^-(-\vec k)] = A[M_u(-\vec k)K^-(\vec k)]= P\cdot
A[M_u(\vec k)K^-(-\vec k)]$$
\pagebreak

{\begin{figure}[htb]
$$\beginpicture
\setcoordinatesystem units <\tdim,\tdim>
\stpltsmbl
\putrule from -25 -30 to 50 -30
\putrule from -25 -30 to -25 30
\putrule from -25 30 to 50 30
\putrule from 50 -30 to 50 30
\plot -25 -20 -50 -20 /
\plot -25 20 -50 20 /
\plot 50 0 120 -20 /
\plot 50 -20 120 -40 /
\photonru 50 20 *3 /
\plot 120 40 90 20 120 20 /
\put {$b$} [b] at -50 25
\put {$\overline{u}$} [t] at -50 -25
\put {$s$} [l] at 125 40
\put {$\overline{u}$} [l] at 125 20
\put {$u$} [l] at 125 -20
\put {$\overline{u}$} [l] at 125 -40
\put {$\Biggr\}$  $K^-(\vec k)$} [l] at 135 30
\put {$\Biggr\}$  $M_u(-\vec k)$}  [l] at 135 -30
\put {$W$} [t] at 70 15
\setshadegrid span <1.5\unitlength>
\hshade -30 -25 50 30 -25 50 /
\linethickness=0pt
\putrule from 0 0 to 0 60
\endpicture$$
\caption{\label{fig-5}} \hfill Color favored tree diagram.
  \hfill~ \end{figure}}

{\begin{figure}[htb]
$$\beginpicture
\setcoordinatesystem units <\tdim,\tdim>
\stpltsmbl
\putrule from -25 -30 to 50 -30
\putrule from -25 -30 to -25 30
\putrule from -25 30 to 50 30
\putrule from 50 -30 to 50 30
\plot -25 -20 -50 -20 /
\plot -25 20 -50 20 /
\plot 50 20 120 40 /
\plot 50 -20 120 -40 /
\photonru 50 0 *3 /
\plot 120 20 90 0 120 -20 /
\put {$b$} [b] at -50 25
\put {$\overline{u}$} [t] at -50 -25
\put {$u$} [l] at 125 40
\put {$\overline{u}$} [l] at 125 20
\put {$s$} [l] at 125 -20
\put {$\overline{u}$} [l] at 125 -40
\put {$\Biggr\}$ $M_u(\vec k)$}  [l] at 135 30
\put {$\Biggr\}$ $K^-(-\vec k)$} [l] at 135 -30
\put {$W$} [t] at 70 -5
\setshadegrid span <1.5\unitlength>
\hshade -30 -25 50 30 -25 50 /
\linethickness=0pt
\putrule from 0 0 to 0 60
\endpicture$$
\caption{\label{fig-6}} \hfill Color suppressed tree diagram.
\hfill~ \end{figure}}

\end{document}